\RequirePackage{fix-cm}
\documentclass[twocolumn]{svjour3}          
\smartqed  
\usepackage{graphicx}
\usepackage{amsmath}
\usepackage{enumerate}
\usepackage{caption}
\usepackage{color}
\usepackage{amsfonts} 
\usepackage{amssymb}
\usepackage{subfigure}
\usepackage{mathptmx}      
\usepackage{latexsym}
\begin{document}

\title{Amplitude modulation of three-dimensional low frequency solitary waves in a magnetized dusty superthermal plasma
}


\author{Shalini \and
        A. P. Misra \and N. S. Saini 
}


\institute{Shalini \at
              Deptt. of Physics, Guru Nanak Dev University, Amritsar \\
              \email{shal.phy29@gmail.com}           
           \and
           A. P. Misra \at
             Deptt. of Mathematics, Siksha Bhavana, Visva-Bharati University, Santiniketan\\
             \email{apmisra@gmail.com}
             \and
           N. S. Saini \at
           Deptt. of Physics, Guru Nanak Dev University, Amritsar\\
             \email{nssaini@yahoo.com}
}
\maketitle

\begin{abstract}
The  amplitude modulation of three dimensional (3D) dust ion-acoustic wave (DIAW) packets is studied in a collisionless magnetized plasma with inertial positive ions, superthermal electrons and negatively charged immobile dust grains. By using the  reductive perturbation technique, a 3D-nonlinear Schr{\"o}dinger (NLS) equation is derived, which governs the slow modulation of DIAW packets. The latter are found to be stable in the low-frequency $(\omega<\omega_c)$ regime, whereas  they are unstable for $\omega>\omega_c$, and the modulational instability (MI) is  related to the modulational obliqueness $(\theta)$.  Here, $\omega~(\omega_c)$ is the nondimensional wave (ion-cyclotron) frequency. It is shown that the superthermal parameter $\kappa$, the frequency $\omega_c$ as well as the charged dust impurity $(0<\mu<1)$ shift the MI domains around the $\omega-\theta$ plane, where $\mu$ is the ratio of electron to ion number densities. Furthermore, it is found that the  decay rate of instability is quenched by the superthermal parameter $\kappa$  with cut-offs at lower wave number of modulation ($K$),  however, it can be higher (lower) with increasing values of $\mu$ ($\omega_c$) having cut-offs at higher values of $K$.
\end{abstract}
 \section{Introduction}\label{sec-intro}
The nonlinear features of  solitary waves in dusty plasmas have been of great importance over the last many years due to their wide range of applications in space, astrophysical and laboratory environments \cite{shukla99,shukla02,boufendi05}. Dust grains are typically micron or sub-micron sized particles and are ubiquitous ingredients in our universe. The presence of charged dust grains in an electron-ion plasmas not only alters the characteristics of ion-acoustic solitary waves (IASWs) but also modifies the ion-acoustic wave  as well as generates a new kind of mode, namely, the dust-acoustic (DA) wave.  More than three decades ago, \cite{shukla92} reported theoretically the existence of DIA waves in a dusty plasma. Later, in  laboratory experiments, \cite{barkan96} confirmed the existence of these waves. A large number of investigations on DIA waves in  multicomponent plasmas have been reported  in the framework of Sagdeev's approach as well as reductive perturbation technique. 
\par
Furthermore, in many   observations it has been  confirmed that superthermal particles exist in space plasmas \cite{summer91,sittler83,mace95,vasyliunas68} and laboratory environments \cite{hellberg00}. These superthermal particles are described by Lorentzian (kappa) distribution which is more appropriate for analysis of data rather than a Maxwellian distribution \cite{vasyliunas68}. Furthermore, such distribution has been widely used to investigate various collective modes as well as nonlinear coherent structures like solitons, shocks, envelope solitons through the description of
Korteweg-de Vries (KdV), Korteweg-de Vries Burgers (KdVB) and nonlinear Schr{\"o}dinger (NLS) equations \cite{saini09,shah11,tantawy11,sultana11,shahmansouri14,adnan14,shahmansouri14a,shalini15}.
\par
On the other hand, there has also been a growing interest in investigating the nonlinear modulation of electrostatic waves in plasmas owing  to their importance not only in space and astrophysical environments but also in laboratory plasmas. The modulational instability (MI) of nonlinear waves in plasmas has been a well-known mechanism for the localization of wave energy, which
leads to the formation of bright envelope solitons. However, in the absence of instability,  the   evolution of the system can be in the form of dark envelope solitons. Furthermore, due to a small plane wave perturbation, MI can have exponential growth which leads to the amplification of
the sidebands, and thus break up the uniform waves into a train of oscillations. A large number of investigations on MI of
electrostatic or electromagnetic waves can be found in the literature  \cite{sultana11,shalini15,kourakis03,kourakis04a,kourakis04b,kourakis04d,misra07,saini08,bains10,taibany06}. To mention few, the MI of obliquely propagating DIA waves in an unmagnetized plasma containing positive ions, electrons and immobile dust grains  was reported by \cite{kourakis03}. It was observed from the stability analysis that the obliqueness in the modulation direction has a profound effect on the condition of MI. They also observed the influence of ion temperature on the amplitude modulation of wave and noticed that wave stability profile may be strongly modified by ion temperature \cite{kourakis04b}. Furthermore,  The nonlinear propagation of wave envelopes   in an unmagnetized superthermal dusty plasma was investigated by El-Labany \cite{labany2017} \textit{et al.} They shown that the electron superthermality  and the dust grain charge significantly modify the profiles of the wave envelope and the  associated  regions of instability.  Ahmadihojatabad  \textit{et al.} \cite{ahmadihojatabad10} studied the influence of superthermal and trapped  electrons  on the obliquely propagating ion-acoustic waves (IAWs) in magnetized  plasmas. Bains \textit{et al.} \cite{bains10} addressed the MI of ion-acoustic wave envelopes in a multicomponent magnetized plasma  using a quantum fluid model. They observed that the ion number density, the constant magnetic field and the quantum coupling parameter have strong effects on the growth rate of MI. Also, the nonlinear propagation of DIA wave envelopes in a three-dimensional magnetized plasma containing nonthermal electrons featuring Tsallis distribution, both positive and negative ions, and immobile charged dust was investigated by \cite{guo14a}.
\par
To the best of our knowledge the investigation of MI of DIAWs in a magnetized dusty plasma containing superthermal electrons has not yet been reported. Our purpose in this investigation is to consider the propagation of  DIA wave envelopes  in a magnetized dusty plasma containing
cold positive ions and superthermal electrons. We have employed the standard multiple-scale perturbation technique to derive the NLS equation. It was shown that in earlier investigations \cite{gharaee11} MI of ion acoustic waves is significantly influenced by the presence of superthermal electrons and growth rate is larger in the presence of more superthermal electrons. We have, however, investigated  the combined effects of the external magnetic field, dust concentration and the superthermality of electrons on the MI of DIA wave packets.
It is shown that the superthermality of electrons (via $\kappa$), the charged dust impurity and the external magnetic field shift the MI domains around the $\omega-\theta$ plane, where $\omega$ is the  wave frequency and $\theta$ stands for modulational obliqueness. Further, we have also studied the decay rate of MI by  different plasma parameters.
\par
The paper is organized as follows: In Sec. \ref{section2}, the basic equations governing the nonlinear dynamics of DIA wave envelopes in magnetized superthermal plasmas are presented and the three-dimensional NLS equation is derived. The effects of various physical parameters on the existence of stable/unstable regions for the modulation of DIA waves are investigated in Sec. \ref{section3}. Finally, Sec. \ref{section4} contains the summary and conclusions of our results.
\section{The Model equations and derivation of the 3D-NLSE}\label{section2}
We consider the nonlinear  propagation of DIA waves in a magnetized plasma consisting of superthermal electrons, cold positive ions and negatively charged immobile dust grains. The plasma is immersed in the constant magnetic field ${\bf B}_{0} = B_{0} \hat{z}$. We adopt a fluid model for the dynamics of DIA waves in a magnetized plasma which consists of the continuity, momentum and the Poisson's equations. Thus, we have

\begin{equation}
\frac{\partial {n}}{\partial {t}}+\nabla\cdot (n{\bf U})=0,\label{em1a}
\end{equation}
\begin{equation}
\frac{\partial {\bf U}}{\partial t}+ ({\bf U}\cdot\nabla){\bf U}=-\frac{e}{M}\nabla \phi+
\frac{eB_0}{M}({\bf U} \times \hat{z}),\label{em2b}
\end{equation}
\begin{equation}
\nabla^2 \phi=4\pi e (n_e-n+Z_dn_{d0}),\label{em3c}
\end{equation}
where the superthermal electrons are given by the   kappa distribution \cite{hellberg09}
\begin{equation}
n_e= n_{e0}\left[1-\frac{e\phi}{(\kappa-3/2)k_B T_e}\right]^{-\kappa+1/2}\label{kappa-dist}.
\end{equation}
The set of fluid equations (\ref{em1a})-(\ref{em3c})  in nondimensional forms  are written as
\begin{equation}
\frac{\partial n}{\partial t}+\nabla\cdot (n{\bf U})=0,\label{em1}
\end{equation}
\begin{equation}
\frac{\partial {\bf U}}{\partial t}+ ({\bf U}\cdot\nabla){\bf U}=\nabla \phi+
\omega_{c}({\bf U} \times \hat{z}),\label{em2}
\end{equation}
\begin{equation}
\nabla^2 \phi=n_e-n+(1-\mu),\label{em3}
\end{equation}
where  $n_e$ and $n$ are the number densities of electrons and ions  normalized by the  equilibrium number density of ions $n_0$,  $\phi$ is the electric
potential normalized by $k_BT_e/e$, ${\bf U}\equiv(u,~v,~ w)$ is the ion fluid velocity normalized by the DIA speed $C_s ~(=\sqrt{k_BT_e/M}$). The space
and time coordinates are normalized by the Debye lengh $\lambda_{D} ~[=\left(k_BT_e/4\pi n_{0}e^2\right)^{1/2}]$ and the inverse of ion plasma frequency $\omega_{pi}~[=\left(4 \pi e^2n_{0}/M \right)^{1/2}]$ respectively. Furthermore, $\omega_{c}=eB_0/cM$ is the ion gyrofrequency normalized by $\omega_{pi}$. The charge neutrality condition yields $1-\mu=Z_d n_{d0}/n_{0}$, where  $\mu=n_{e0}/n_{0}$ is the ratio of equilibrium number densities of electrons and ions. Next, in the small-amplitude perturbations, i.e., $|\phi|\ll1$, and, in particular, $|\phi/(\kappa-3/2)|\ll1$, Eq. (\ref{kappa-dist}) reduces to
\begin{equation}
n_e\approx \mu+q_1\phi+q_2\phi^2+q_3\phi^3, \label{ne-kappa}
\end{equation}
where the coefficients are given by
\begin{equation}
\begin{split}
&q_1=\mu\frac{(\kappa-1/2)}{(\kappa-3/2)},~ q_2=\mu\frac{(\kappa^2-1/4)}{2(\kappa-3/2)^2},\\ &q_3=\mu\frac{(\kappa^2-1/4)(\kappa+3/2)}{6(\kappa-3/2)^3}.\label{em4}
\end{split}
\end{equation}
In order to derive the evolution equation for weakly nonlinear DIA wave envelopes, we employ the standard multiple scale technique
\cite{taniuti69, asano69} in which the coordinates are stretched as
\begin{equation}
\xi=\epsilon x,~\eta=\epsilon y,~\zeta=\epsilon\left(z-V_gt\right),~\tau=\epsilon^2 t.\label{stretch}
\end{equation}
Consider $A\equiv(n,~w,~\phi)$ and $B\equiv(u,~v)$ as the state vectors which describe the state at a position $z$ and time $t$. The perturbations from the equilibrium state $A^{(0)}=(1,0,0)^T$ and $B^{(0)}=(0,0)^{T'}$ are considered by assuming $A=A^{(0)}+\sum_{m=1}^{\infty}\epsilon^mA^{(m)}$ and $B=B^{(0)}+\sum_{m=1}^{\infty}\epsilon^{m+1}B^{(m)}$. The slow-scale dependence of all perturbed state enter via the $l$-th harmonic amplitude $A_l^{(m)}$ and $B_l^{(m)}$ given as $A^{(m)}=\sum_{l=-m}^{m}A_{l}^{(m)}(\xi,\eta,\zeta,\tau)e^{il(kz-\omega t)}$ and $B^{(m)}=\sum_{l=-m}^{m}B_{l}^{(m)}(\xi,\eta,\zeta,\tau)e^{il(kz-\omega t)}$, where $\omega$ and $k$, respectively, represent the carrier wave frequency and the wavenumber. In order that  $n$, ${\bf U}$, $\phi$ etc. are all real, the state variables must satisfy the reality condition with respect to its complex conjugate parts.  One should note that the transverse (to the magnetic field) velocity components $u$ and $v$  appear at higher order in $\epsilon$ than the parallel component $w$. The anisotropy and higher order effects are introduced via strong magnetic field and gyro-motion of fluid respectively in the presence of weak perturbations \cite{xue05}.
\par
\begin{figure*}[htb!]
\centering
\includegraphics[height=3.2in, width=6.5in]{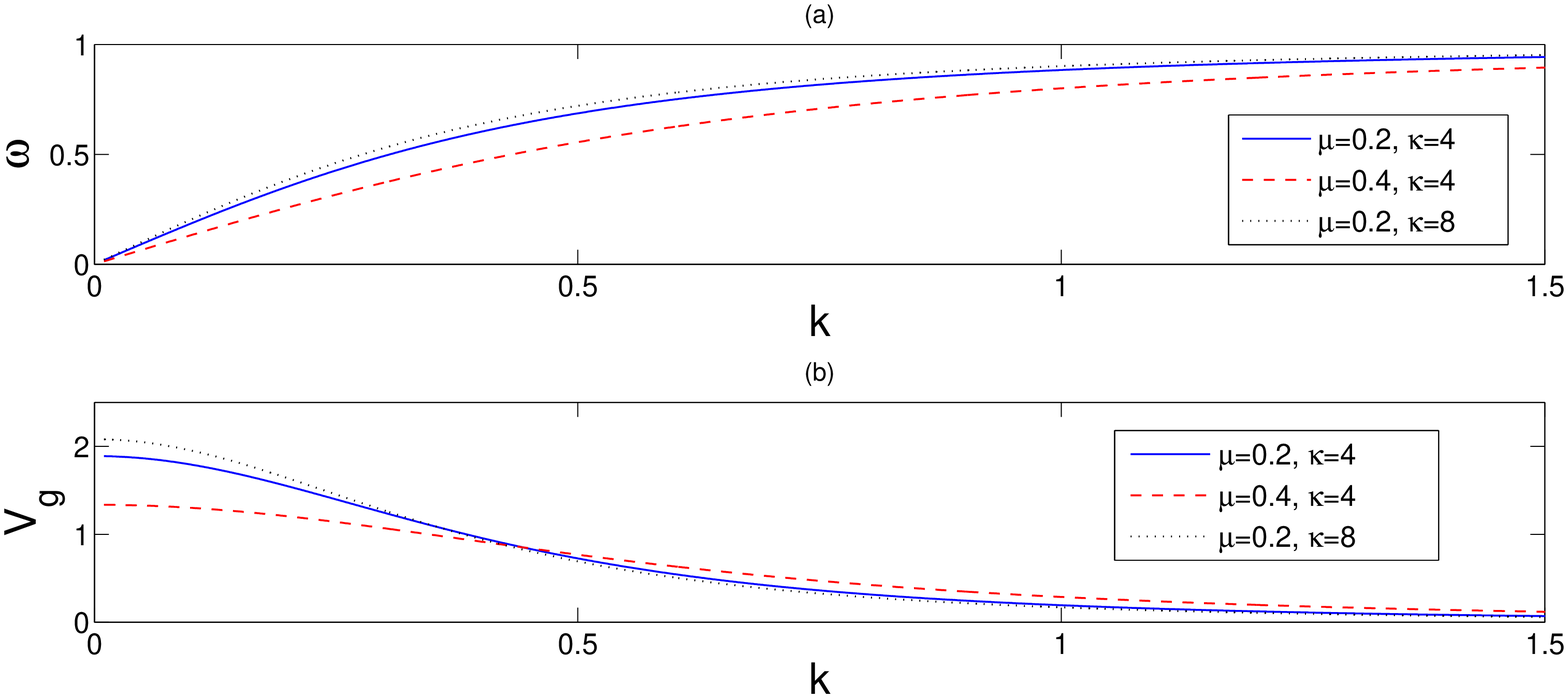}
\caption{Carrier wave frequency $\omega$ (upper panel) and the group velocity of the wave packet $V_g$ (lower panel) are plotted against the wave number $k$ for different values of $\kappa$ and $\mu$ as shown in the legends.}
\label{fig13d}
\end{figure*}

We substitute the stretched coordinates  (\ref{stretch}) and the expansions given above into Eqs. (\ref{em1})-(\ref{ne-kappa}), and collect  terms in   different powers of $\epsilon$  to   obtain a set of reduced equations. Thus, equating the coefficients for $m=1$ and  $l=1$, we obtain the following first-order quantities in terms of $\phi_1^{(1)}$
\begin{equation}
n_{1}^{(1)}=\frac{k^2}{\omega^2}\phi_1^{(1)},\label{em6}
\end{equation}
\begin{equation}
w_{1}^{(1)}=\frac{k}{\omega}\phi_1^{(1)},\label{em7}
\end{equation}
together with the linear dispersion relation
\begin{equation}
\omega^2=\frac{k^2}{k^2+ q_1}.\label{em8-3d}
\end{equation}
From the second order reduced equations ($m=2$, $l=1$), the following compatibility condition in terms of the group velocity of waves is obtained as
\begin{equation}
V_g\equiv\frac{\partial \omega}{\partial k}=q_1\frac{\omega^3}{k^3}.\label{em9-3d}
\end{equation}
We have depicted the variation of the carrier wave frequency [Eq. \ref{em8-3d}] and the group velocity [Eq. \ref{em9-3d}] against the carrier wave number $k$ in Fig. \ref{fig13d}. From the upper panel, we find that as $k$ increases, the frequency $\omega$ increases and it approaches
a constant value at higher $k~(>1)$. Furthermore, $\omega$ (normalized by the ion plasma frequency $\omega_{pi}$) increases and approaches a constant value (close to $1$)  as the wave number $k$ increases. Also, the value of $\omega$ increases with increasing values of the   spectral index $\kappa$ (i.e., when the superthermality of electrons is somewhat relaxed), however,  it remains almost unaltered for    $\kappa>8$. Nevertheless, a reduction of the wave frequency is noticed with increasing values of the electron to ion number density ratio $\mu$. This implies that as the number density of electrons increases, more electrons will flow out of the dust grains, i.e., dust charge number decreases in order to maintain the quasineutrality. Further increase of $\mu$ may eventually lead to the case similar to the dust free electron-ion plasma. Thus, negatively charged dust impurity in the plasma with $n_{0}>n_{e0}$ effectively increases the wave frequency. Such  dust impurity has also a significant effect on the group velocity of waves as shown in the lower panel of Fig. \ref{fig13d} (see the solid and dashed lines). The group velocity of waves ($V_g$) decreases with an increase in the wave number ($k$) for different values of $\kappa$ and $\mu$. It is very interesting to see that for smaller $k<0.5$ (i.e, for larger wavelength), the group velocity reduces with an increase in $\mu$, however,    it increases with larger $k$. Furthermore, an increase in the parameter $\kappa$ (e.g., from $\kappa=4$ to $\kappa=8$) leads to an enhancement of $V_g$ as $k\rightarrow0$. Here, note that further increase of $\kappa~(>8)$ does not give any significant change in $V_g$. 

For $l=0$, $1$ and $2$, we can determine the second order harmonic modes in terms of $\phi_1^{(1)}$. So, for $m=2$, $l=1$, we have
the   reduced  equations
\begin{equation}
i \omega n_{1}^{(2)}+i k w_{1}^{(2)}=V_g\frac{\partial n_{1}^{(1)}}{\partial \zeta},\label{em10}
\end{equation}
\begin{equation}
i \omega w_{1}^{(2)}+i k \phi_{1}^{(2)}=V_g\frac{\partial w_{1}^{(1)}}{\partial \zeta}-\frac{\partial \phi_{1}^{(1)}}{\partial \zeta},
\label{em11}
\end{equation}
with
\begin{equation}
u_{1}^{(1)}=\frac{\omega_{c}\frac{\partial \phi_{1}^{(1)}}{\partial \eta}-i \omega \frac{\partial \phi_{1}^{(1)}}{\partial \xi}}{\omega^2-\omega_{c}^2},\label{em12}
\end{equation}

\begin{equation}
v_{1}^{(1)}=-\left[\frac{i\omega\frac{\partial \phi_{1}^{(1)}}{\partial \eta}+\omega_{c} \frac{\partial \phi_{1}^{(1)}}{\partial \xi}}
{\omega^2-\omega_{c}^2}\right].\label{em13}
\end{equation}
The second order harmonic modes with $m=2$ and $l=2$   are  given by
\begin{equation}
-\omega n_{2}^{(2)}+k w_{2}^{(2)}+kn_{1}^{(1)}w_{1}^{(1)}=0,\label{em14}
\end{equation}
\begin{equation}
-2\omega w_{2}^{(2)}+k (w_{1}^{(1)})^2+k\phi_{2}^{(2)}=0.\label{em15}
\end{equation}
Thus, we  obtain
\begin{equation}
n_{2}^{(2)}=C_1^{(22)}(\phi_{1}^{(1)})^2,~u_{2}^{(2)}=C_2^{(22)}(\phi_{1}^{(1)})^2,\label{em16}
\end{equation}
\begin{equation}
\phi_{2}^{(2)}=C_3^{(22)}(\phi_{1}^{(1)})^2,\label{em17}
\end{equation}
where the coefficients are
\begin{equation}
C_1^{(22)}=(4k^2+q_1)C_3^{(22)}+q_2,\label{em18}
\end{equation}
\begin{equation}
C_2^{(22)}=\frac{\omega}{k}\left[C_1^{(22)}-(k^2+q_1)^2\right],\label{em19}
\end{equation}
\begin{equation}
C_3^{(22)}=\frac{q_2}{3 k^2}+\frac{k^2}{2\omega^4}.\label{em20}
\end{equation}
We note that    the first order zeroth harmonics ($n_{0}^{(1)}$,~ $w_{0}^{(1)}$,~ $\phi_{0}^{(1)}$) vanish \cite{taniuti74}, which gives $u_{0}^{(1)}=v_{0}^{(1)}=0$. For $m=2$, $l=0$, we   obtain  the second order and zeroth order harmonic modes in the following forms
\begin{equation}
n_{0}^{(2)}=C_{1}^{(20)}\left(\phi_{1}^{(1)}\right)^{2},~u_{0}^{(2)}=C_{2}^{(20)}\left(\phi_{1}^{(1)}\right)^{2},
\label{em21a}
\end{equation}
\begin{equation}
\phi_{0}^{(2)}=C_{3}^{(20)}\left(\phi_{1}^{(1)}\right)^{2},\label{em21}
\end{equation}
where the coefficients are
\begin{equation}
C_1^{(20)}=q_1C_3^{(20)}+2q_2,\label{em22}
\end{equation}
\begin{equation}
C_2^{(20)}=V_gC_1^{(20)}-2\frac{\omega}{k}(k^2+q_1)^2,\label{em23}
\end{equation}
\begin{equation}
C_3^{(20)}=\frac{2q_2V_g^2-(k^2+3q_1)}{1-q_1 V_g^2}.\label{em24}
\end{equation}
Proceeding to the next order ($m=3$) and solving for the first harmonic equations ($l=1$), an explicit compatibility condition is determined, from which we   obtain the following    NLS  equation  for  $\Phi\equiv\phi_1^{(1)}$
\begin{equation}
i\frac{\partial \Phi}{\partial \tau}+P\frac{\partial^{2}\Phi}{\partial \zeta^{2}}+Q|\Phi|^{2}\Phi-S\left(\frac{\partial^2\Phi}{\partial \xi^2}+
\frac{\partial^2\Phi}{\partial\eta^2}\right)=0.\label{em7}
\end{equation}
The coefficient of dispersion $P$ and the nonlinearity $Q$ are given by
\begin{equation}
P\equiv\omega''(k)=-\frac{3}{2}q_1\frac{\omega^5}{k^4},\label{em8}
\end{equation}
\begin{equation}
\begin{split}
Q=&\frac{\omega^3}{k^2}\left[\frac{3}{2} q_3+ q_2\{C_3^{(20)}+C_3^{(22)}\}\right]\\
&-\frac{\omega}{2}\{C_1^{(20)}+C_1^{(22)}\}-k\{C_2^{(22)}+C_2^{(20)}\}.\label{em9}
\end{split}
\end{equation}
The coefficient $S$ which accounts for the combined effects of transverse perturbations and the external magnetic field is given by
\begin{equation}
S=\frac{\omega^3}{2k^2(\omega_{c}^2-\omega^2)}.\label{em10}
\end{equation}
\section{Stability analysis}\label{section3}
We note that the amplitude modulation of DIA wave envelopes typically depend on the coefficients of the NLS equation (\ref{em7}), which
parametrically depend on the density ratio $\mu$, the superthermality of electrons (via $\kappa$) as well as the intensity of the magnetic field (via $\omega_{c}$). Inspecting the coefficients $P,~Q$ and $S$, we find that $P=-(3/2) q_1\frac{\omega^5}{k^4}\equiv -(3/2)\frac{\omega^3}{k^2}(1-\omega^2)$, i.e., $P$ is always negative for $\kappa>3/2$ (for which $q_1>0$) and $\omega<1$. However, $Q$ can be positive or negative depending on the values of $k,~\mu$ and $\kappa$. Also, $S>0~(S<0)$ according to when $\omega<\omega_c~(\omega>\omega_c)$. We will find that the key elements responsible for the MI are the ratios $P/Q$ and $S/P$ together with their signs and  magnitudes. Considering a harmonic wave solution of Eq. (\ref{em7}) of the form $\Phi=\Phi_{0}\exp(iQ|\Phi_{0}|^{2}\tau)$ with $\Phi_{0}$ denoting the constant amplitude, one can obtain the following dispersion relation for the modulated DIA wave packets \cite{bains10}.
\begin{eqnarray}
\Omega^{2}=&&K^{4}\left(\frac{P\alpha^{2}-S}{1+\alpha^{2}}\right)^{2}\nonumber\\
&&\times\left(1-\frac{2(1+\alpha^{2})|\Phi_{0}|^{2}}{K^{2}}\frac{Q/P}{\alpha^{2}
-S/P}\right),\label{em11}
\end{eqnarray}
where $\Omega$ and $K\equiv\sqrt{K_{\xi}^{2}+K_{\eta}^{2}+K_{\zeta}^{2}}$, respectively, denote the wave frequency and the wave number of
modulation. The parameter $\alpha\equiv K_{\zeta}/\sqrt{K_{\xi}^{2}+K_{\eta}^{2}}$ is related to the modulational obliqueness   $\theta$ which the wave vector $\mathbf{K}$ makes with the resultant of $K_{\xi}\hat{x}$ and $K_{\eta}\hat{y}$, i.e., $\theta=\arctan(\alpha)$.
From Eq. (\ref{em11}), we find that there exists a critical wave number $K_{c}$ such that $K^{2}<K_{c}^{2}\equiv2|\Phi_{0}|^{2}(1+\alpha^{2})
(Q/P)/(\alpha^{2}-S/P)$, the MI sets in either for $PQ>0$, $\alpha_1^{2}-S/P>0$ or for $PQ<0$, $\alpha_1^{2}-S/P<0$ \cite{bains10}. It is further found that a critical value of $\theta$, i.e.,  $\theta_{c}\equiv \arctan(\sqrt{S/P})$ also exists for the occurrence of $MI$. Thus, the MI may occur either for $PQ>0$, $\theta>\theta_{c}$ or $PQ<0$, $\theta<\theta_{c}$, i.e.,  we have two possible cases:
\begin{itemize}
\item Case I: When $\omega<\omega_c$, the MI sets in for $Q<0$ and for any value of $\theta$ in $0\leq\theta\lesssim\pi/2$.
\item {Case II: When $\omega>\omega_c$, the MI sets in either for $Q<0$ and  $\theta>\theta_c$ or $Q>0$ and  $\theta<\theta_c$.}
\end{itemize}
\par
\begin{figure*}[htb!]
\includegraphics[height=3.5in, width=6.5in]{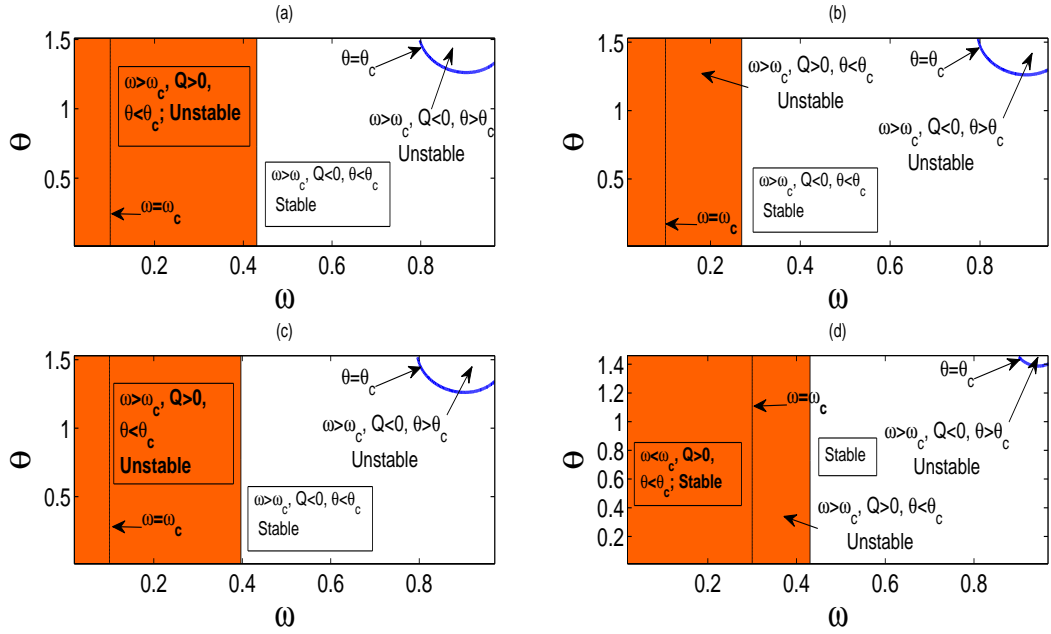}
\caption{The stable and unstable regions of wave modulation  are shown by the contour plots of $Q=0,~\omega=\omega_c$ and $\theta=\theta_c$ in the $\omega-\theta$ plane for different values of the parameters: (a)  $\kappa=4,~\omega_c=0.1$ and $\mu=0.2$, (b)  $\kappa=4,~\omega_c=0.1$ and $\mu=0.3$, (c) $\kappa=6,~\omega_c=0.1$ and $\mu=0.2$, and (d) $\kappa=4,~\omega_c=0.3$ and $\mu=0.2$.   The shaded or gray (blank or white) region stands for  $Q>0~(Q<0)$. When $\omega>\omega_c$, the $MI$ occurs either for $Q>0,~\theta<\theta_c$, or for $Q<0,~\theta>\theta_c$. No instability occurs in the regime $\omega<\omega_c$.    In some other regions the modulated wave becomes stable. }
\label{fig2}
\end{figure*}

From the subsequent analysis and Fig. \ref{fig2} it will be clear that  the Case I is not admissible to the present study as there is no common region for which $\omega<\omega_c$ and   $Q<0$ are satisfied. So, we will focus only on Case II. It turns out that when the DIA wave frequency is larger than the ion-cyclotron frequency $\omega_{c}$, the MI is related to the  obliqueness parameter $\theta$, however, the instability disappears for $\omega<\omega_c$.  We numerically investigate different stable and unstable regions in the $\omega-\theta$ plane as shown in Fig. \ref{fig2}. We find that the charged dust impurity (represented by the parameter $\mu$ with $0<\mu<1$), the superthermal parameter $\kappa$ and the gyrofrequency $\omega_c$ shift the stable/unstable regions around the $\omega-\theta$ plane.   From panels (a) and (b) it is clear that as $\mu$ increases, i.e., as the number of charged dust grains decreases, a part of the  instability  region (with $\theta<\theta_c$)  shifts to a stable one and the region of stability in the $\omega-\theta$ plane increases. However,   the instability region with $\theta>\theta_c$ increases slightly with increasing values of $\mu$. This implies that when the obliqueness parameter $\theta$ is below its critical value $\theta_c$, the presence of charged dust impurity in the plasma favors the instability of modulated wave packets. Comparing panel (c) with panel (a) we find that the superthermality of electrons (with lower values of $\kappa$)  also favors the instability in the  region with $\theta<\theta_c$. The instability region with $\theta>\theta_c$ remains almost unchanged.     From panels (a) and (d) it is also evident that  the external magnetic field   significantly reduces the regions of instability both  in  the cases of $\theta<\theta_c$ and $\theta>\theta_c$.

\begin{figure*}[htb!]
\centering
\includegraphics[height=3.2in, width=6.5in]{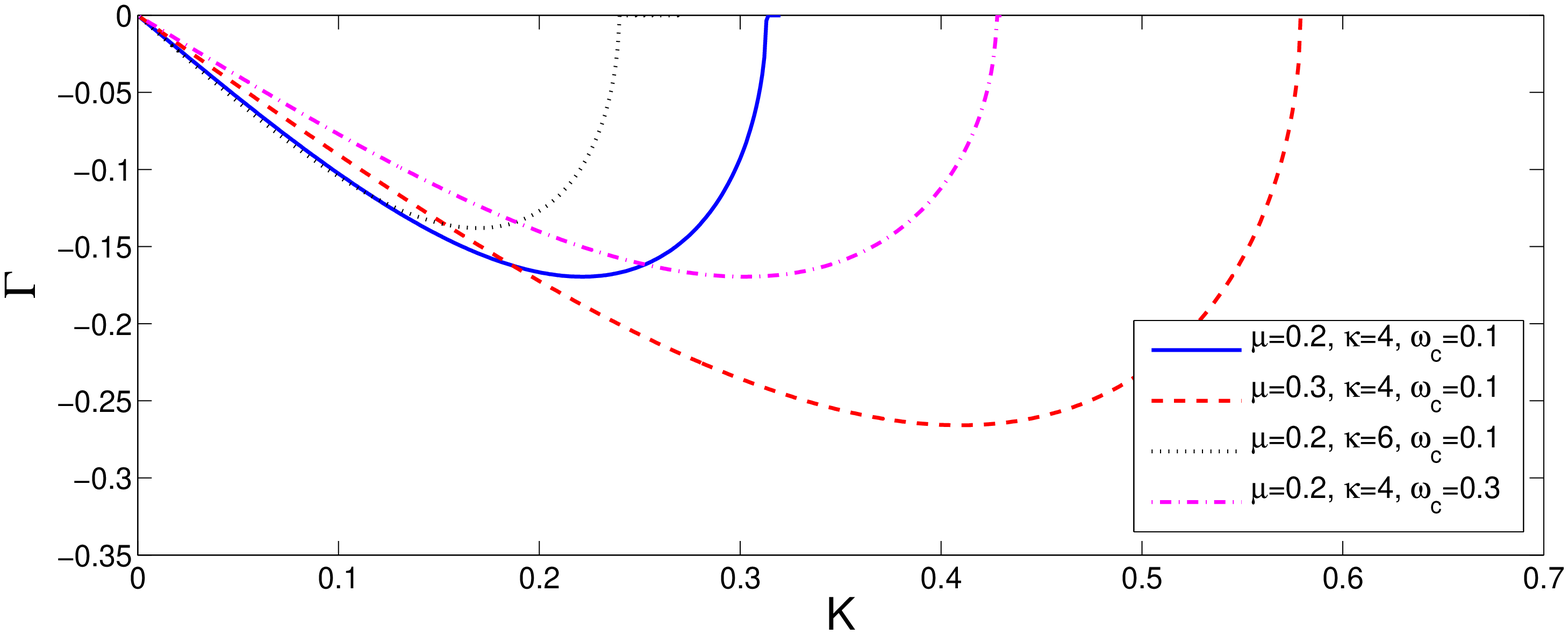}
\caption{The decay rate of modulational instability $\Gamma$ is shown against the wave number of modulation $K$ with the variation of the  parameters as shown in the legend. The other parameter values are $k=0.5$ and $\phi_0=0.5$.}
\label{fig3}
\end{figure*}

The maximum growth/decay rate $\Gamma_{max}$=Im $\left(\Omega\right)_{max}$ can be obtained from Eq. (\ref{em11}) as
$\Gamma_{max}=Q|\phi_{0}|^{2}$ provided $PK_{\zeta}^{2}-S(K_{\xi}^{2}+K_{\eta}^{2})=Q|\phi_{0}|^{2}$ is satisfied. The decay rate of MI is depicted in Fig. \ref{fig3} for different values of $\kappa$, $\mu$ and $\omega_c$.
Clearly, the effects of higher values of $\kappa$ (less superthermality) suppresses the instability decay rate with cutoffs at significantly lower wave numbers of modulation (see the solid and dotted lines). However, the decay rate becomes higher with increasing values of the  electron concentration (or decreasing the dust concentration)  with cutoffs at higher  $K$ (see the solid and dashed lines). We find that in contrast to the unmagnetized plasmas, the effect of $\omega_c$ is to increase the cutoffs at higher wave numbers of modulation, however, the decay rate is slightly reduced (see the solid and dash-dotted lines).

\section{Summary and Conclusion}\label{section4}

We have investigated the amplitude modulation of DIA wave packets in a magnetized multi-component plasma consisting of singly charged positive
ions, superthermal electrons featuring kappa distribution and negatively charged immobile dust grains. Using the multiple scale
technique, a NLS equation is derived which governs the evolution of DIA wave envelopes.  It is shown that both the dispersive and the nonlinear coefficients of the NLS equation are significantly modified by  the effects of charged dust impurity,    the external magnetic field as well as the superthermality of electrons. Different stable and unstable regions under modulation are obtained in the plane of the carrier
wave frequency ($\omega$) and the obliqueness ($\theta$) of modulation. It is found that the parameters $\kappa$, $\mu$, $\omega_c$ and $\theta$ remarkably shift the stable/unstable regions around the $\omega-\theta$ plane. The growth/decay rate of instability is also examined numerically with these plasma parameters. The main results are summarized as follows:

\begin{itemize}
\item Starting from a set of fluid equations, the dynamics of weakly nonlinear, slowly varying DIA wave packets is shown to be governed by
 a  three-dimensional NLS equation in which the additional dispersive terms (leading to two more space dimensions in the equation) appear due to the combined effects of the transverse perturbations and the external magnetic field.  The fluid model with $\kappa$-distributed electrons and stationary charged dust particles is valid for the plasma parameters satisfying $0<\mu<1$ and $\kappa>3/2$.

\item The carrier wave frequency is seen to assume a constant value at large $k>1$, and approaches the ion plasma frequency with increasing
  values of $\kappa$. The wave frequency $\omega$ and hence the group velocity $V_g$ get significantly  reduced with higher values of $\mu$.

\item The group velocity dispersion $P$ of the NLS  equation is always negative irrespective of the values of $k$ and the plasma parameters. The nonlinear coefficient $Q$ is always negative for $k\gtrsim 1$, however, it can be either positive or negative   in the range $0<k<1$ depending on the values of $\kappa$ and $\mu$. For  propagation below the ion cyclotron frequency, the DIA wave packet is always stable. However, for  $\omega>\omega_c$,  it is unstable and the MI is  related to the obliqueness parameter $\theta$. The parameters $\kappa$, $\mu$ and $\omega_c$ are found to shift the instability regions around the $\omega-\theta$ plane significantly.

\item The decay rate of MI is found to be significantly suppressed  by the  effects of $\kappa$, i.e., when  $\kappa$ increases with cutoffs at lower wave   numbers of modulation. However, it can be higher with increasing values of the density ratio $\mu$. The effect of the external magnetic field is to decrease the decay rate with   cutoffs  at higher values of the wave number of modulation.
\end{itemize}
\par
The findings of the present investigation may be useful for the modulation of dust-ion acoustic wave envelopes in dusty superthermal plasmas such as those in laboratory \cite{lab},   space \cite{space} and astrophysical \cite{astro} environments.

\acknowledgement
Shalini thanks University Grants Commission, New Delhi for awarding Rajiv-Gandhi Fellowship. A. P. M acknowledges support from UGC-SAP (DRS, Phase III) with  Sanction  order No.  F.510/3/DRS-III/2015(SAPI) dated  25/03/2015, and UGC-MRP with F. No. 43-539/2014 (SR) and FD Diary No. 3668 dated 17.09.2015. The work of N.S.S. was supported by University Grants Commission, New Delhi, India under the major research project $F. No. 41-873/2012(SR)$.


\begin{thebibliography}{100}
\bibitem{shukla99}Shukla P. K., Mendis D. A. and Desai T., \textit{Advances in Dusty Plasmas}, World Scientific, Singapore (1999).
\bibitem{shukla02}Shukla P. K. and Mamun A. A., \textit{Introduction to Dusty Plasma Physics}, Institute of Physics, Bristol (2002).
\bibitem{boufendi05}Boufendi L., Mikikian M. and Shukla P. K., \textit{New Vistas in Dusty Plasmas}, AIP Proceeding, AIP, New York (2005).
\bibitem{shukla92}Shukla P. K. and Silin V. P., \textit{Physica Scripta}, \textbf{45}, 508 (1992).
\bibitem{barkan96}Barkan A., D'Angelo N. and Merlino R., \textit{Planet. Space. Sci.}, \textbf{44}, 239 (1996).
\bibitem{summer91}Summers D. and Thorne R. M., \textit{Phys. Fluids B}, \textbf{3}, 1835 (1991).
\bibitem{sittler83}Sittler Jr. E. C., Ogilvie K. W. and Scudder J. D., \textit{J. Geophys. Res.}, \textbf{88}, 8847 (1983).
\bibitem{mace95}Mace R. L. and Hellberg M. A., \textit{Phys. Plasmas}, \textbf{2}, 2098 (1995).
\bibitem{vasyliunas68}Vasyliunas V. M., \textit{J. Geophys. Res.}, \textbf{73}, 2839 (1968).
\bibitem{hellberg00}Hellberg M. A., Mace R. L., Armstrong R. J. and Karlstad G., \textit{J. Plasma Phys.}, \textbf{64}, 433 (2000).
\bibitem{saini09}Saini N. S., Kourakis I. and  Hellberg M. A., \textit{Phys. Plasmas}, \textbf{16},
    062903 (2009).
\bibitem{shah11}Shah A., Mahmood S. and Haque Q., \textit{Phys. Plasmas}, \textbf{18}, 114501 (2011).
\bibitem{tantawy11}El-Tantawy S. A., El-Bedwehy N. A. and Moslem W. M. \textit{Phys. Plasmas}, \textbf{18}, 052113 (2011).
\bibitem{sultana11}Sultana S. and Kourakis I., \textit{Plasma Phys. Control. Fusion} \textbf{53}, 045003 (2011).
\bibitem{shahmansouri14} Shahmansouri M. and Tribeche M.,  \textit{Astrophys. Space Sci.}, \textbf{350}, 045003 (2014).
\bibitem{adnan14}Adnan M., Mahmood S. and Qamar A., \textit{Adv. Space Res.}, \textbf{53}, 845 (2014).
\bibitem{shahmansouri14a} Shahmansouri, M. and  Astaraki E.,  \textit{J. Theor. Appl. Phys.}, 8, 189 (2014).
\bibitem{shalini15} Shalini, Saini N. S. and Misra A. P. \textit{Phys. Plasmas}, \textbf{22}, 092124 (2015).
\bibitem{kourakis03}Kourakis I. and Shukla P. K., \textit{J. Phys. A: Math. Gen.} \textbf{36}, 11901 (2003).
\bibitem{kourakis04a}Kourakis I. and Shukla P. K. \textit{Phys. Rev. E.}, \textbf{69}, 036411 (2004).
\bibitem{kourakis04b}Kourakis I. and Shukla P. K. \textit{Euro. Phys. J. D.}, \textbf{28}, 109 (2004).
\bibitem{kourakis04d} Kourakis I. and Shukla P. K. \textit{Phys. Scr.}, \textbf{69}, 316 (2004).
\bibitem{misra07}Misra A. P. and Bhowmik C. \textit{Phys. Plasmas}, \textbf{14}, 012309 (2007).
\bibitem{saini08}Saini N. S. and Kourakis I. \textit{Phys. Plasmas}, \textbf{15}, 123701 (2008).
\bibitem{bains10}Bains A. S., Misra A. P., Saini N. S. and Gill T. S., \textit{Phys. Plasmas}, \textbf{17}, 012103 (2010).
\bibitem{taibany06}El-Taibany W. F. and Kourakis I., \textit{Phys. Plasmas}, \textbf{13}, 062302 (2006).
\bibitem{labany2017} El-Labany S. K., El-Shewy  E. K.,  Abd El-Razek H. N. and  El-Rahman A. A.,  \textit{Adv. Space Research},  \textbf{59},  1962  (2017).
\bibitem{ahmadihojatabad10}Ahmadihojatabad N., Abbasi H. and Hakimi Pajouh H., \textit{Phys. Plasmas}, \textbf{17}, 112305 (2010).
\bibitem{guo14a}Guo Shimin and Mei Liquan, \textit{Phys. Plasmas}, \textbf{21}, 82303 (2014).
\bibitem{gharaee11}Gharaee H., Afghah S. and Abbasi H. \textit{Phys. Plasmas}, \textbf{18}, 032116 (2011).
\bibitem{hellberg09}Hellberg M. A., Mace R. L., Baluku T. K., Kourakis I. and Saini N. S., \textit{Phys. Plasmas} \textbf{16}, 094701 (2009).
\bibitem{taniuti69}Taniuti T. and Yajima N., \textit{J. Math. Phys.}, \textbf{10}, 1369 (1969).
\bibitem{asano69}Asano N., Taniuti T. and Yajima N., \textit{J. Math. Phys.}, \textbf{10}, 2020 (1969).
\bibitem{xue05}Xue Ju-Kui, \textit{Phys. Plasmas}, \textbf{12}, 062313 (2005).
\bibitem{taniuti74}Taniuti T., \textit{Progress of Theoretical Physics Supplement}, \textbf{55}, 1 (1974).
\bibitem{lab}  Liu J. M.,  DeGroot J. S.,  Matte J. P.,  Johnston T. W. and 
Drake R. P., \textit{Phys. Rev. Lett.}, \textbf{72}, 2717 (1994).
\bibitem{space}  Montgomery M. D.,  Bame S. J.  and  Hundhausen A. J., \textit{J. Geophys. Res.}, \textbf{73}, 4999 (1968); Maksimovic M.,  Pierrard V. and  Riley P., \textit{Geophys. Res. Lett.}, \textbf{24}, 1151 (1997);  Zouganelis I., \textit{J. Geophys. Res.}, \textbf{113}, A08111 (2008).
\bibitem{astro}  Pierrard V. and  Lazar M., \textit{Solar Phys.}, \textbf{267}, 153 (2010), and
references therein.                            
\end{thebibliography}
\end{document}